\documentclass[12pt,preprint]{aastex}
\def\t0{\theta_{\circ}}

\def\be{\begin{equation}}
\def\en{\end{equation}}

\def\msun{M_{\sun}}

\def\lsun{L_{\sun}}

\newcommand{\um}{$\mu$m}                          %micron
\newcommand{\asecdot}[2]{\mbox{#1$\stackrel {\prime \prime}{_{\bf \cdot}}$#2}}

\begin{document}

\shorttitle{Companions to HD 199143 and HD 358623}
\shortauthors{Jayawardhana \& Brandeker}

\title {Discovery of close companions to the nearby young stars\\
 HD 199143 and HD 358623\altaffilmark{1}}
\author{Ray Jayawardhana}
\affil{Department of Astronomy, University of California, Berkeley, CA 94720, U.S.A.}

\email{rayjay@astro.berkeley.edu}

\and

\author{Alexis Brandeker}
\affil{Stockholm Observatory, SCFAB, SE-106 91 Stockholm, Sweden}

\email{alexis@astro.su.se}

\altaffiltext{1}{Based on observations collected at the European Southern Observatory, Chile.}

\begin{abstract}
Young stellar systems in the solar neighborhood provide valuable laboratories 
for detailed studies of star and planet formation. The bright F8V star 
HD 199143 and the Li-rich late-type emission line star HD 358623 are among 
the nearest young stars identified to date, and may be members of a young 
association in Capricornus. We present high-resolution near-infrared images 
of these two sources, obtained using the adaptive optics system on the 
3.6-meter telescope at the European Southern Observatory in La Silla, Chile. 
Our observations reveal that both are in fact close binary systems. The newly 
discovered companion at a separation of $\sim$1'' may account for the unusual 
characteristics of HD 199143 --rapid rotation, emission lines, ultraviolet 
variability, and excess infrared emission-- recently discussed by van den 
Ancker and co-workers. HD 199143 may be a rare example of a close binary with 
only a circum{\it secondary} disk. With the detection of a $\sim$2'' 
companion, HD 358623 is now possibly one of the closest known T Tauri 
binaries. Both binary systems are prime targets for follow-up spectroscopic 
and astrometric observations. 
\end{abstract}


\section{Introduction}
There is growing interest in finding and characterizing young stellar
systems in the solar neighborhood since their proximity offers unique 
advantages for improving our understanding of star and planet formation. 
Over the past five years, a number of young associations have been 
identified within 100 pc of the Sun, and these have become the targets
of intense investigations using a variety of techniques (Jayawardhana \& 
Greene 2001).

HD 199143 is a bright F8V (Houk \& Smith-Moore 1988) star in Capricornus
with a {\it Hipparcos} distance of 47.7 $\pm$ 2.4 pc. It was detected as a 
bright extreme-ultraviolet source by the {\it ROSAT} and {\it Extreme 
Ultraviolet Explorer} satellites (Pounds et al. 1993; Malina et al. 1994). 
Recently, in a study of its optical and ultraviolet spectra, van den Ancker 
et al. (2000) found the emission lines of Mg II, C I, C II, C III, C IV, Si 
IV, He II, and N V and a high level of variability, both in the continuum and 
line fluxes. The authors also present evidence for very rapid (a few hundred 
km s$^{-1}$) rotation of the stellar photosphere, and for excess emission in 
the mid-infrared, detected by {\it IRAS} at 12$\mu$m. Van den Ancker et al. 
proposed that all of these phenomena can be explained if HD 199143 has
a low-mass chromospherically active companion that dominates the ultraviolet
and infrared emission of the system. Episodic accretion of material from a 
putative T Tauri like companion could have spun up the primary and may also 
account for the bursting or flaring nature of this object.

The same authors pointed out that the photometrically variable K7-M0e
dwarf HD 358623 (BD-17 6128), previously studied by Mathioudakis et al. 
(1995), is located only 5 arcminutes from HD 199143. The optical spectrum of 
HD 358623 is identical to that of many classical T Tauri stars, and shows 
strong H$\alpha$ emission and Li 6708 \AA~ absorption lines. The closeness 
of the two stars on the
sky and the similarity of their proper motions, derived from the Tycho-2
catalog, led van den Ancker et al. (2000) to suggest that the pair 
belong to a physical group that may include additional young stars. At a 
distance of $\sim$48 pc, HD 358623 is estimated to be $\sim10^7$ years old 
--an age comparable to the timescales of inner disk evolution and planet 
formation (Jayawardhana 2001 and references therein).

Given their proximity, interesting age and possible membership in a co-moving
stellar group, HD 199143 and HD 358623 are worthy of further study. Here we 
present the results of near-infrared adaptive optics imaging observations of 
both stars that reveal each to be a close binary. The newly discovered
companion to HD 199143 may indeed account for the engimatic characteristics
of that system.

\section{Observations}
We observed HD\,199143 and HD\,358623 on 2001 May 31 to June 1 with the
European Southern Observatory 3.6-meter telescope at La Silla, Chile, using
the adaptive optics near infrared system (ADONIS) and the SHARPII+ camera. The
SHARPII+ camera is based on a 256$\times$256 NICMOS~III array. We used a
plate scale of 0.035\,arcsec/pixel, Nyquist-sampling the diffraction limited
point spread function (PSF) in $J$ and giving an effective field of view of
\asecdot{8}{5}$\times$\asecdot{8}{5}. We observed the targets by taking a
series of frames on source and chopping out for sky frames. For HD\,199143
we obtained 42 frames in $J$, each of integration time 3\,s, and 42 frames
in $K$ of 2\,s each. For HD\,358623 we obtained 10$\times$20\,s in $J$ and
22$\times$10\,s in $K$. The science target itself was used as the wavefront
sensor of the adaptive optics (AO) system. The seeing during the observations,
as reported by the La Silla seeing monitor, was about \asecdot{1}{3} in
$V$ (0.55\,\um).

After each series of observations of the science targets, an identical series
of observations with the same AO correction parameters were obtained of a
PSF calibrator star, for post processing purposes. For HD\,199143 we used
the F3IV star SAO\,163887 with $V$ magnitude 6.8, and for HD\,358623 we
used the M0 star SAO\,163911 with $V$ magnitude 9.8.

\section{Data Reduction and Analysis}
The basic data reductions were done in a standard way by subtracting sky
frames from source frames and then dividing by a flat field obtained on
sky during dusk. The resulting frames were then not stacked together but
instead processed with the myopic deconvolution algorithm
IDAC\footnote{Information about IDAC and the package itself may be found
on the web pages of ESO, {\it http://www.eso.org/}}.

 From the PSF calibrator star observations, we obtained an initial guess of
the PSF during the science data acquisition. Since the PSF calibrator star
data were not obtained simultaneously with the science data, the initial
guess PSF is then modified by IDAC iteratively to better match the
different frames of science data. A series of PSF estimates corresponding
to the science object series is thus produced along with the deconvolved
image. The deconvolution procedure using IDAC took about two weeks of
processing time on an UltraSparc workstation. We refer to Christou et al.
(1999) for details on IDAC.

We performed astrometry of the HD\,199143 and HD\,358623 systems using the
deconvolved images in both $J$ and $K$. The error in separation was estimated
by checking the consistency of measured separations in $J$ and $K$.
For flux calibration, we used observations of the IR standard star
HD\,190285 and 5 other SAO stars in both $J$ and $K$. For the SAO stars,
 known $B$ and $V$ magnitudes and spectral classes were used to derive their
$J$ and $K$ magnitudes using standard colors (Bessell \& Brett 1988) and an
estimate of the color excess due to extinction using Table~1 of Mathis
(1990), where we assumed an optical extinction ratio $R_V=3.1$. We then
fitted a relation between the observed count rates and derived magnitudes.

 From the fit we estimated the absolute photometric errors, i.e. the
uncertainty of the transformation from measured counts to magnitudes.
By checking the consistency of the measured flux of stars observed several
times during the night we derived the relative photometric errors.

\section{Results and Discussion}
We have resolved both HD 199143 and HD 358623 into close binaries for the first
time. Figure 1 shows the deconvolved $J$ and $K$ images, and Table 1 presents 
photometric and astrometric measurements of each system. Our observations are 
able to detect binaries at separations between \asecdot{0}{08} and $4''$, 
where the lower limit corresponds to the diffraction limit of the 3.6-m 
telescope in $J$, and the upper limit is due to the 
\asecdot{8}{5}$\times$\asecdot{8}{5} field of view. A search of the 2MASS
catalog reveals that the surface density of stars brighter than $K=9$ 
within a 5-degree radius of our targets is $\sim$0.004 arcmin$^{-2}$. Thus, it
is unlikely that the companions we found are unrelated field stars.

Our $J-K$ color, as well as the $B-V$ color from the SIMBAD database, of
HD 199143 A is consistent that of a F8V star with no foreground extinction.
Using a distance of 48 pc and the atmosphere models of Allard, Hauschildt
\& Schweitzer (2000), we find a {\it bolometric} luminosity of $L_{bol, A}=
2.45\pm0.12\lsun$. Stellar evolutionary models of Palla \& Stahler (1999)
yield a mass of $M_A=1.19\pm0.02\msun$ and an age of 18$\pm$2 Myrs for the
primary. (The errors given for the stellar parameters are the formal 
uncertainties in statistical model fitting plus measurement errors; ``true'' 
errors --due to inherent uncertainties in the models --are likely to
be larger.)

The newly discovered companion, HD 199143 B, is extremely red, with
$J-K$=1.368. This near-infrared color corresponds to a model atmosphere with
a temperature in the 1700-2100 K range, which (taken at face value) would 
imply a very young object in the brown dwarf/giant planet mass regime. 
However, HD 199143 B most likely harbors a circumstellar disk, responsible for 
$K$-band excess as well as the IRAS 12$\mu$m detection and ground-based
mid-infrared measurements previously reported for the system (van den Ancker
et al. 2001). Given the lack of optical colors and a spectral type, it
is difficult to derive reliable stellar parameters for the secondary. However,
assuming an age of 18 Myrs and the excess to be zero at $J$ and a free 
parameter at $K$, we find $L_{bol, B}=0.060\pm0.003\lsun$, 
$M_B=0.35\pm0.05\msun$, and a $J-K$ excess of $\sim$0.5 mag, consistent with 
a T Tauri like star. 

If HD 199143 B indeed harbors a dusty disk but HD 199143 A does not, a
scenario in agreement with our data, it would be a rare example of a close 
binary with only a circum{\it secondary} disk. In a study of 25 close,  
pre-main-sequence binaries, Prato (1998) found disks around both stars in 
15 systems and neither star in 5 systems. Four others showed evidence of
circum{\it primary} disks (also see Jayawardhana et al. 1999) while only one 
example of a circum{\it secondary} disk was identified. 

Theoretical calculations suggest that circumstellar disks will be truncated
by the tidal effects of a companion star in circular orbit at approximately 
0.9 of the average Roche lobe radius (Artymowicz \& Lubow 1994). From the
{\it IRAS} database, van den Ancker et al. (2000) derived a flux of 
0.24$\pm$0.04 Jy in the 12$\mu$m band, and upper limits of 0.12, 0.12, and 
0.30 Jy for the fluxes at 25, 60 and 100$\mu$m, respectively. A tidally
truncated disk around HD 199143 B is consistent with the tight {\it IRAS}
limits at 25 and 60$\mu$m.

Our $J-K$ color and the SIMBAD $B-V$ color for HD 358623 A imply very little
extinction ($\le$0.1mag). For a K7 spectral type, assuming an age of 18 Myrs
to be coeval with HD 199143, we find $L_{bol, A}=0.21\pm0.04\lsun$ and 
$M_A=0.74\pm0.07\msun$ for a distance of 43 pc. Conversely, if we fix the
distance at 48 pc, we get a slightly worse fit to the evolutionary models
with an age of 12$\pm$4 Myrs, again in reasonable agreement with the age of
HD 199143. Using a distance of 43 pc, stellar models yield 
$L_{bol, B}=0.034\pm0.002\lsun$, $M_B=0.27\pm0.06\msun$ and an age of 
20$\pm$7 Myrs for the newly found secondary, HD 358623 B. Van den Ancker et
al. (2001) report significant mid-infrared excesses from HD 358623. Our 
observations do not constrain whether the dust surrounds one or both members 
of the binary.

\section{Conclusions}
Using adaptive optics techniques, we have resolved the two nearby ($\sim$48 
pc) young stars HD 199143 and HD 358623 into close binaries. The newly 
discovered secondary may account for many of the puzzling characteristics
of the HD 199143 system. If the B component harbors a dusty disk, as 
indicated by its near-infrared colors and consistent with previous 
mid-infrared detections,
HD 199143 would be a rare example of a close binary with only a 
circum{\it secondary} disk. The stellar parameters we derive for each
component of HD 199143 and HD 358623, while not tightly constrained, are
in agreement with both systems being 15$\pm$5 Myrs in age at a distance
of 43-48 pc. Spatially resolved spectroscopy of both binaries would allow
us to determine the stellar parameters more reliably, and follow-up 
astrometry could yield direct dynamical masses, thus testing evolutionary
models of pre-main-sequence stars. High-resolution mid-infrared imaging will 
be valuable for further constraining the location and nature of circumstellar 
dust in these two binaries. A systematic search for other young stars in
the vicinity of HD 199143 and HD 358623 may reveal additional members of
a co-moving stellar group, enhancing our understanding of local star formation
and enriching our inventory of suitable nearby targets for detailed 
multi-wavelength studies.

\bigskip
We wish to thank the staff of the European Southern Observatory for their 
outstanding support. We are grateful to Tom Greene for a previous attempt
to observe HD 199143 at the IRTF, and to the referee, Mario van den Ancker,
for a prompt review. Our research made use of the ADS, SIMBAD and 2MASS 
databases. RJ holds a Miller Research Fellowship at the University of 
California, Berkeley. This work was supported in part by a NASA grant to RJ
administered by the AAS.

%\newpage

\clearpage
\begin{figure}
%\plotfiddle{mosaic2.ps}{4in}{0.}{100.}{100.}{-320}{-250}
\plotone{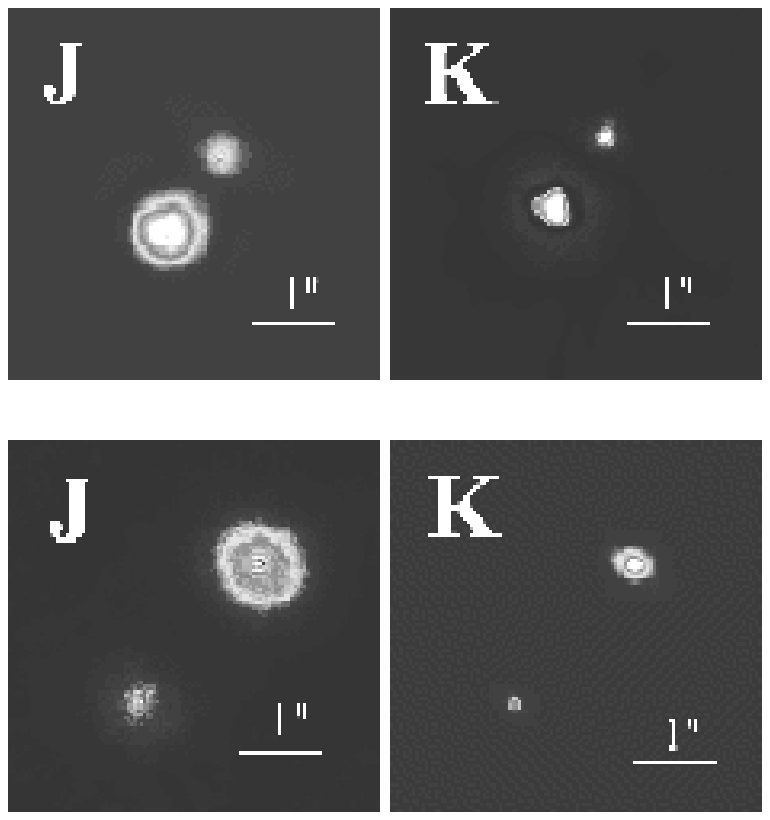}
\caption{Deconvolved adaptive optics images of HD 199143 (top) and HD 358623 
(bottom) in the near-infrared. The horizontal bar in each image corresponds 
to one arcsecond. North is up, east is to the left.
}
%\label{fig:mosaic}
\end{figure}

\clearpage
\begin{table}
\begin{scriptsize}
\begin{center}
\renewcommand{\arraystretch}{1.2}
\begin{tabular}{lccccc}
\multicolumn{6}{c}{\scriptsize TABLE 1}\\
\multicolumn{6}{c}{\scriptsize }\\
\multicolumn{6}{c}{\scriptsize INFRARED PHOTOMETRY AND ASTROMETRY OF HD 199143 AND HD 358623}\\
\multicolumn{6}{c}{\scriptsize }\\
\hline
\hline
%\vspace{0.2cm}
Stellar & [$J$] 1.25\,\um\tablenotemark{a} & [$K$] 2.18\,\um\tablenotemark{b} & $J-K$ & Separation & Position Angle\tablenotemark{c} \\
component & (mag) & (mag) & (mag) & ($''$) & ($\arcdeg$) \\
\hline
HD 199143 A+B & 6.1816 & 5.7725 & ...   & ...  & ... \\
HD 199143 A   & 6.2346 & 5.9054 & 0.329 & ...  & ...  \\
HD 199143 B   & 9.4869 & 8.1191 & 1.368 & $1.0785\pm0.009$ & $325.0\pm0.7$ \\
\hline
HD 358623 A+B & 7.8332 & 7.0071 & ...   & ...  & ...  \\
HD 358623 A   & 8.0201 & 7.2171 & 0.803 & ...  & ...  \\
HD 358623 B   & 9.8359 & 8.8941 & 0.942 & $2.2033\pm0.007$ & $139.4\pm0.5$ \\
\hline
\multicolumn{6}{c}{\scriptsize }\\
\multicolumn{6}{l}{\scriptsize $^a$ Absolute error $\sigma_{J}$=0.041mag, mean relative error $\sigma_{J}$=0.006}\\
\multicolumn{6}{l}{\scriptsize $^b$ Absolute error $\sigma_{K}$=0.032mag, mean relative error $\sigma_{K}$=0.023}\\
\multicolumn{6}{l}{\scriptsize $^c$ Position angle is measured from north to east, relative to A.}
\end{tabular}
\end{center}
\end{scriptsize}
\end{table}

\end{document}